\documentclass{vgtc}                          

\graphicspath{{figures/}{pictures/}{images/}{./}} 

\usepackage{times}                     

\usepackage{tabu}                      
\usepackage{booktabs}                  
\usepackage{lipsum}                    
\usepackage{mwe}                       

\usepackage{mathptmx}                  

\onlineid{0}

\vgtccategory{Research}

\vgtcinsertpkg

\title{Toward a Scientific Discovery Engine for Weather and Climate Data: A Visual Analytics Workbench for Embedding-Based Exploration }

\author{Nihanth W. Cherukuru\thanks{e-mail: ncheruku@ucar.edu} %
\and Matt Rehme\thanks{e-mail: mrehme@ucar.edu} %
\and Kirsten J. Mayer\thanks{e-mail: kjmayer@ucar.edu} %
\and David John Gagne\thanks{e-mail: dgagne@ucar.edu} %
\and John Schreck\thanks{e-mail: schreck@ucar.edu} %
\and John Clyne\thanks{e-mail: clyne@ucar.edu}
\and Charlie Becker\thanks{e-mail: cbecker@ucar.edu}}
\affiliation{\scriptsize NSF National Center for Atmospheric Research, Boulder, Colorado USA}

\teaser{
  \centering
  \includegraphics[width=\linewidth]{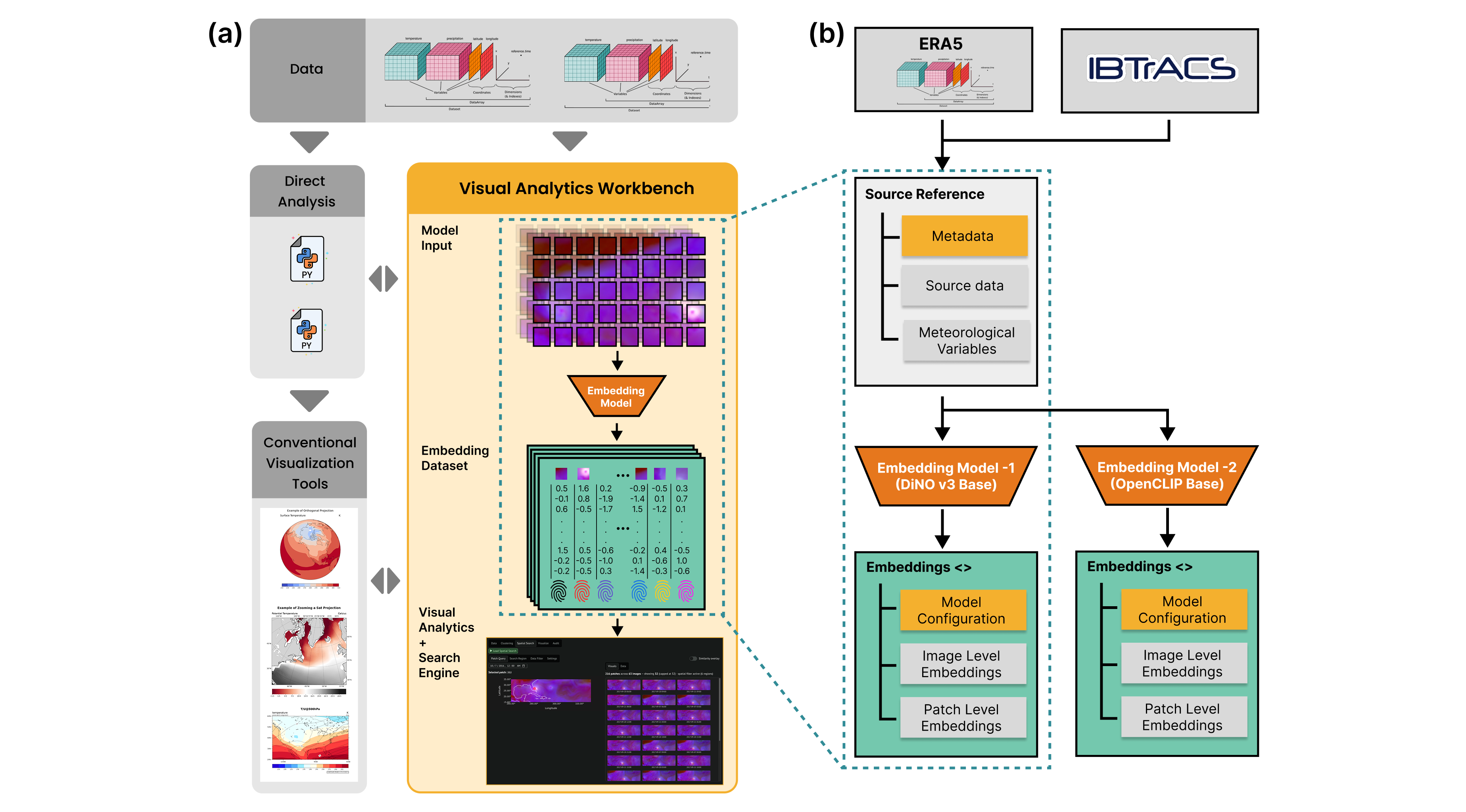}
  \caption{ Overview of the workbench and example retrieval workflow with vision transformer based embedding models. (a) The visual analytics workbench augments conventional weather and climate analysis (shown in gray) by supporting a workflow for generating image and patch level embeddings and subsequent interactive exploration. (b) The system separates a shared source reference (In this example- RGB inputs) containing metadata, source imagery, and meteorological variables, from model-specific embedding experiments (shown in green). Each experiment stores its model configuration and searchable embedding representations, enabling multiple models to be examined in the visual interface over the same source data}
  \label{fig:teaser}
}

\abstract{
Earth system science is producing increasingly large, high-dimensional datasets from both physics-based and AI-driven models. While embedding-based representations make these data searchable and serve as foundational building blocks for AI-driven discovery engines, nearest neighbors in latent spaces are not automatically scientifically meaningful. They may reflect real meteorological structures, or simply artifacts of preprocessing, geography, or model bias. Researchers therefore need visual tools to inspect latent space organization, trace search results back to physical evidence, and evaluate candidate representations against one another.

We present an open source visual analytics workbench designed to support this provenance-aware scientific retrieval workflow. The system links distinct embedding experiments to shared source data, metadata, spatial contexts, and model configurations. It enables interactive retrieval strategy design by allowing users to issue image-level and localized patch-level queries, apply multi-constraint filters, and inspect analogs through familiar meteorological views. This facilitates a discovery loop where scientists characterize a phenomenon in a well-understood dataset and use its latent signature to probe larger archives. While we demonstrate the workbench through a tropical cyclone retrieval scenario using a vision foundation model (DINOv3) on ERA5 data, the framework is model-agnostic and designed to integrate with other embedding architectures in the future. Finally, we evaluate its out-of-core retrieval backend, demonstrating that interactive visual search over tens of millions of embeddings is highly scalable on commodity hardware.
} 


\begin{document}

\firstsection{Introduction}
\maketitle
Numerical weather prediction and Earth system models already produce petabyte-scale, multivariate, spatiotemporal datasets. AI-based weather and climate models compound this: they enable far larger ensembles and cheaper experimentation, opening the door to broader exploration of rare and extreme events. As dataset size grows and the field moves toward combining machine-learning methods with physics-based simulation, the bottleneck increasingly shifts from technical feasibility of producing forecasts to searching, interpreting, and validating the resulting data.

Embedding-based representations offer a promising way to make these datasets searchable \cite{brown_alphaearth_2025}. By mapping complex atmospheric states, spatial regions, or model outputs into vector spaces, embeddings support similarity search, analog retrieval, clustering, and discovery of recurring patterns. This is attractive for researchers in Earth systems science, where there is often a need to find structurally related events across large archives or ensembles. Furthermore, embeddings serve as a natural bridge to integrate scientific data with large language models (LLMs) and agentic workflows. By providing a unified, machine-interpretable representation of complex multimodal data, embeddings enable AI agents to reason over, retrieve, and potentially synthesize scientific knowledge forming a foundational building block for the next generation of AI-driven scientific research and discovery engines \cite{ding_bridging_2026}.

However, embedding-based search functionality might not be inherently scientifically meaningful. A nearest neighbor in latent space may reflect a physically relevant weather structure. It may also reflect artifacts of preprocessing, geography, seasonality, model bias, or the particular representation used to generate the embeddings. This raises a trustworthiness challenge for scientific search: researchers need to determine whether retrieved neighbors are meaningful enough to support scientific reasoning, hypothesis generation, or downstream analysis. Beyond a search interface, users must be able to inspect how embedding spaces are organized, compare alternative representations, formulate retrieval strategies, and verify results against physical meteorological evidence. In practice, these steps are difficult to combine. Embedding generation, dimensionality reduction, vector search, model auditing, geospatial inspection, and conventional meteorological visualization are often handled by separate scripts or tools. This fragmentation makes it hard to determine which embedding models and query strategies are appropriate for a given scientific task.

As an initial step toward addressing this overarching vision, we present an early-stage prototype of an open-source visual analytics workbench designed to demonstrate the core workflows required for developing and auditing embedding-based search for weather and climate data. The system provides a conceptual framework and practical mechanism to link embeddings from different experiments to shared source data, metadata, spatial context, model configuration, and retrieval parameters, allowing users to inspect latent-space organization and verify retrieved neighbors against meteorological evidence. This paper makes three contributions: (a) \textbf{A provenance-aware scientific retrieval workflow}. The workbench connects embedding vectors, source imagery, spatial coordinates, timestamps, metadata, and model configurations so that latent-space results can be traced back to physical data that produced it. This also allows users to test whether nearest neighbors in the latent space reflect meaningful atmospheric structures rather than representation artifacts (b) \textbf{Interactive retrieval-strategy design for meteorological analog search}. The system supports both image-level and localized patch-level visual search, letting users query on whole scenes or on specific features within datasets. The underlying data architecture links the latent space to metadata, spatial coordinates, and user-defined variables, so users can formulate complex, multi-constraint queries to find analogs across datasets (c) \textbf{A demonstration of scalable retrieval and latent exploration}. We utilize a tropical-cyclone retrieval scenario and a vision foundation model (DINOv3) primarily as a vehicle to showcase how embeddings from a particular model can be leveraged for interactive search, retrieval, and latent space exploration. We evaluate the disk-backed vector search backend over tens of millions of high-dimensional embeddings, demonstrating the technical feasibility of scaling this visual workflow across large datasets at interactive speeds.

The source code, example data, and instructions for the demonstrated workflows are released as supplemental material (\cref{sec:supplemental_materials}) to support reproducibility and extension.

\section{Background}

\subsection{Big Data Challenges in Earth System Modeling}

Navigating petabyte-scale climate data was identified as a major challenge more than a decade ago \cite{overpeck_climate_2011} and remains a research priority \cite{sellars_grand_2018, vance_big_2024}. Recent advances in AI-based Earth system modeling further intensify this challenge. AI weather prediction models now outperform operational numerical weather prediction systems on several metrics while requiring substantially fewer computational resources \cite{bouallegue_rise_2024}. This also enables much larger ensembles \cite{ mahesh_huge_2025}, which are important for uncertainty quantification, rare-event analysis, and characterizing variability in chaotic atmospheric systems. As Earth system models increasingly combine machine learning with physical modeling \cite{eyring_pushing_2024}, analysis workflows must support not only access to high-volume outputs, but also their interpretation, validation, and search.
    
Substantial work has addressed the technical challenge of accessing and visualizing large Earth science datasets. Multi-resolution progressive access \cite{panta_web-based_2024}, cloud-native chunking strategies \cite{abernathey_cloud-native_2021}, and scientific visualization tools remain essential for moving from raw data to verifiable visual evidence. However, access and rendering alone do not solve the problem of deciding what to inspect. Visual analytics systems address this gap by combining automated analysis with interactive exploration. Keim et al. \cite{keim_visual_2008} argue for an “analyze first” workflow in which systems identify potentially important structure before users visually inspect details. Our work follows this tradition but focuses on embedding-based search as the automated analysis layer for weather and climate data, while preserving links to meteorological evidence for interpretation and validation.

\subsection{Embeddings, Search, and Latent Space Exploration Tools}

Representation learning is increasingly used to identify structure in Earth science data \cite{liu_representation_2025}. Climate research has a long history of vector-space representations, including empirical orthogonal functions based on principal component analysis \cite{monahan_empirical_2009}; modern embeddings extend this idea using nonlinear architectures such as autoencoders, contrastive models, and foundation models. Recent examples include autoencoder-based ERA5 representations \cite{zhao_transforming_2025}, AlphaEarth Foundations and its embedding dataset for Earth observation \cite{brown_alphaearth_2025}, analyses of Aurora model embeddings \cite{richards_latent_2026}, and mechanistic interpretation tools for weather and climate model latent spaces \cite{tempest_mechanistic_2026}.

Embedding-based search complements supervised event detection methods such as ClimateNet \cite{prabhat_climatenet_2021} by enabling users to query for similar atmospheric states or localized structures without relying only on predefined labels. This is important for discovery tasks involving rare events, hidden precursors, or “unknown unknowns,” where unsupervised and self-supervised methods can reduce dependence on expert labels \cite{molina_review_2023}. Several recent systems address adjacent pieces of this problem. Embedding Atlas \cite{ren_embedding_2025} provides a low-friction, general-purpose interface for inspecting large embedding projections through scalable rendering, density-based clustering, and metadata cross-filtering, with the projection itself as the primary analysis surface. ClimateSOM \cite{kawakami_climatesom_2026} targets climate ensembles, using a self-organizing map and LLM-assisted sensemaking to compare variability across ensemble runs by laying members onto a learned 2D grid. 

Our workbench is designed for a different approach: embedding inspection and similarity retrieval are coupled stages of a single workflow. The system supports multiple coexisting embedding experiments over the same source data, with provenance back to the underlying meteorology, and the retrieval backend is built for out-of-core search over patch-level indexes that exceed workstation memory. The pace of representation learning also matters for system design. Different models, preprocessing choices, spatial scales, and similarity definitions yield different vector spaces and different retrieval behavior, and the right representation for a given scientific question is rarely known in advance. We therefore frame embedding-based retrieval as a visual analytics problem in which model comparison, search, interpretation, and scalable access are coupled requirements, and we organize the system around an embedding experiment abstraction that lets multiple representations of the same source data coexist and be evaluated against each other.

\section{Design Goals}

Embedding-based search is promising for exploring large weather and climate datasets, but it is difficult to use scientifically unless users can understand what is being searched, how similarity is defined and whether retrieved results correspond to meaningful physical phenomena. Drawing on prior work and conversations with domain scientists, we designed the workbench around three goals:

\noindent\textbf{G1: Support provenance-aware embedding experiments.}
Each representation (embedding dataset) is treated as an experiment tied to a shared source dataset, capturing model configuration, preprocessing, spatial sampling, timestamps, metadata, and source imagery. This lets researchers load and sequentially compare representations from different architectures (E.g. vision models, autoencoders, domain foundation models, latent spaces from AI weather emulators), within the same visual analytics interface, while tracing any latent-space result back to the data and modeling choices that produced it.

\noindent\textbf{G2: Support interactive, domain-grounded retrieval design.}
What counts as meteorological similarity depends on spatial scale, metadata, geographic context, and the query formulation. Users should therefore be able to build and revise retrieval strategies (E.g. global-level, localized patch-level, metadata-constrained, spatially constrained, and staged global-to-local) and inspect results through familiar domain views (E.g. maps, timestamps, metadata, source imagery, derived fields, and highlighted matching regions). Together these let users judge whether nearest neighbors reflect real atmospheric structure.

\noindent\textbf{G3: Support scalable out-of-core retrieval.}
Embedding workflows can produce many vectors per timestep, ensemble member, variable, or region, quickly exceeding workstation memory. The system must support interactive retrieval at this scale without requiring all vectors in main memory or persistent access to high-memory HPC environments. This matters most during experimentation, when several embedding collections may coexist for the same dataset.

\section{System Design And Workflow}
At a high level, the workbench links two representations of the same data within the same interface: the physical meteorological view, consisting of visualizations of raw multi-variate, spatio-temporal data; and the latent space of vectors produced by an embedding model (\cref{fig:teaser}(a)). Since the conventional meteorological visualization capabilities follow standard practice, we do not detail them here and instead demonstrate them in the supplemental video.
The interactive workflow operates over pre-computed embeddings. Embeddings are produced through an embedding model applied to the source data. Because embedding generation could be a resource-intensive, asynchronous batch process that scales with model complexity and dataset size, it is deliberately decoupled from the interactive visual workflow. This design keeps the interface responsive during exploration, and treating generation as a separate process lets new embedding models be added without changing the interactive workflow. The released source code includes options to generate embeddings from two vision models: DINOv3 and OpenCLIP. The interactive workflow then proceeds as an iterative loop. The user visualizes the data, selects a patch or region of interest, retrieves the embedding of that query, and fetches and displays the most similar analogs, which can be traced back to their source data.
\subsection{Data Architecture}
The platform's central abstraction is an embedding experiment: a set of vectors generated from a shared source reference using an embedding model, preprocessing pipeline, spatial sampling scheme, and configuration (G1, \cref{fig:teaser}(b) - shown as green boxes). This abstraction keeps the system model-agnostic. Because experiments are defined against a shared source reference, the same source data can support multiple embedding models, each stored as an independent experiment with its own configuration and image and patch-level embeddings \cref{fig:teaser}(b). The source agnostic architecture enables future extensions to support and index vectors from any model that produces a fixed-length vector representation such as pretrained foundation models, custom autoencoders, or domain-specific encoders, as long as the resulting vectors can be linked back to their source data and metadata.

During the exploratory visual analysis stage, users first load an experiment and inspect its associated source metadata and the embedding model configuration. This inspection step lets users verify which dataset, representation, and spatial sampling scheme they are analyzing before using the embedding space for search. The workbench supports two complementary forms of interaction. First, users can examine the organization of an embedding space through dimensionality-reduction views such as PCA and UMAP. These projections can be linked with metadata brushing, parallel coordinates, image galleries, and domain-specific views, allowing users to inspect whether latent-space structure corresponds to time, location, event labels, or other physical attributes.

Second, users can construct search queries and evaluate retrieval strategies. A query may represent an entire atmospheric state, a localized spatial region, or a staged combination of both (G2). Retrieval can be constrained by metadata or spatial filters, and the returned analogs are linked back to source data, timestamps, similarity scores, and highlighted matching regions when available. This makes search an iterative visual process: users can adjust the query formulation and constraints, inspect the retrieved analog, and decide whether the strategy captures the intended meteorological similarity.

\subsection{Retrieval Mechanism}
For any given embedding experiment, the workbench holds linked data from the source reference table and the embedding experiment tables: the embedding vectors (for vision transformers, both image-level and patch-level vectors), any user-defined variables, and spatial and temporal tags. Retrieval is the process of finding the items whose embeddings are most similar to a query. For a patch-level search, the user selects a query patch, and the system compares that patch's query vector against the other patch embedding vectors, ranking them by a similarity score. The similarity score is a scalar measure of closeness in the latent space, where a higher score indicates greater semantic similarity. The items with the highest scores are returned as the nearest analogs. Beyond similarity alone, the user-defined variables, and spatial and temporal tags provide additional ways to filter the search, restricting retrieval to items that match chosen metadata values or fall within a specified region or time range.

Executing this search at interactive speed over large collections raises two challenges: returning the nearest neighbors with low latency, and storing them compactly enough that the index need not be held entirely in memory. This is necessary since extracting patch-level representations multiplies the vector count by the number of spatial tokens per image for each timestep, ensemble member, or variable. The backend addresses these challenges through disk-backed vector indexes in Lance \cite{pace_lance_2025} (G3) and employing inverted-file (IVF) indexing and product quantization (PQ) described below.

\begin{table}[tb]
  \caption{Three-channel encoding used to render ERA5 fields as RGB
  inputs to DINOv3. This mapping exists to satisfy the model's
  three-channel input requirement and to let users inspect what the
  model sees. It is a diagnostic aid, not a substitute for the
  workbench's conventional meteorological visualizations.}
  \label{tab:rgb-encoding}
  \scriptsize%
  \centering%
  \begin{tabu}{%
    l%
    l%
    l%
    l%
    }
  \toprule
  Channel & Variable & Range (Unit) & Scaling \\
  \midrule
  Red   & MSL Pressure Anomaly     & $[-20, 20]$ hPa          & Inverted \\
  Green & 10\,m Wind Speed         & $[0, 35]$ m\,s$^{-1}$    & Linear \\
  Blue  & Total Column Water Vapor & $[20, 70]$ kg\,m$^{-2}$  & Square-root \\
  \bottomrule
  \end{tabu}%
\end{table}
\begin{figure}[tb]
 \centering 
 \includegraphics[width=\columnwidth]{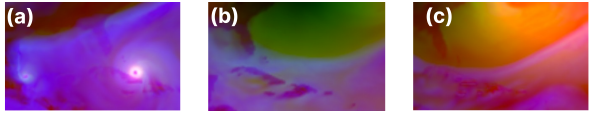}
 \caption{Example composites showing the RGB encoding and how it
highlights different weather regimes used as input to DINOv3:
(a) tropical cyclones with compact bright centers, (b) cold arctic
air intrusion with green-dominant wind and low-moisture patterns,
and (c) extratropical cyclones with broader synoptic structure.}
 \label{fig:composite}
 \end{figure}

After vector embeddings are generated from the dataset, Inverted File (IVF) indexing clusters these vectors into partitions, each represented by a centroid. During the search and retrieval phase, the algorithm probes only these centroids of the partitions nearest to the query rather than scanning the entire dataset, reducing the number of vectors examined per query and thus accelerating interactive retrieval. Combined with disk-backed, memory-mapped storage, this allows only the relevant partitions to be loaded on demand rather than holding the full index in RAM, keeping the runtime memory footprint low \cite{johnson_billion-scale_2021}. Because the underlying vision model outputs L2-normalized embeddings, ranking by cosine similarity is equivalent to ranking by squared Euclidean distance. Cosine similarity is also the conventional metric for vision embeddings, whose semantic structure is carried in vector orientation rather than magnitude. When a query includes metadata or spatial constraints, these can be applied as either a prefilter, restricting the nearest-neighbor search to valid rows, or a postfilter, applied to the search results, with the choice exposed to the user per query. Additionally, Product Quantization (PQ) splits high-dimensional floating-point vectors into subvectors and stores them as integer codes, substantially compressing the memory footprint occupied by the raw embeddings \cite{jegou_product_2011}. PQ is a form of lossy compression which is viable in this application since the workflow relies on semantic similarity and not exact mathematical reconstruction. Because it offers large memory savings with minimal impact on approximate nearest-neighbor search quality, PQ is a standard optimization widely adopted by modern vector databases. We note that PQ is not a required feature for the system to function. Users may instead store raw embeddings directly, or reduce dimensionality using PCA or similar techniques during their own preprocessing before importing embeddings into the workbench, depending on their memory constraints and fidelity requirements. \cref{sec_scalability} provides a scalability analysis of this retrieval backend.

\begin{figure}[t]
  \centering
  \includegraphics[width=\columnwidth]{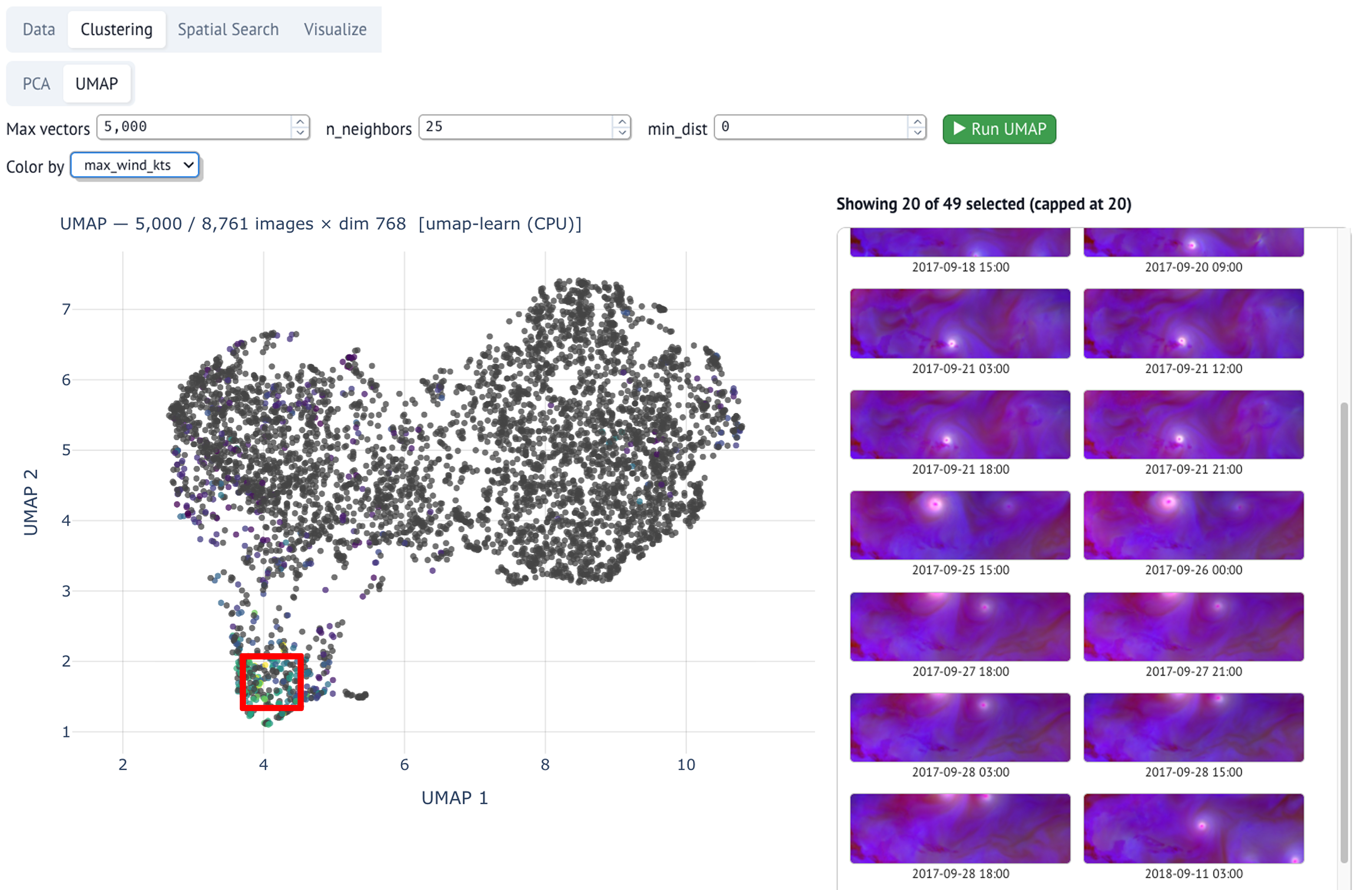}
  \caption{The UMAP view projects image-level embeddings colored by maximum IBTrACS wind speed; selecting a high-wind cluster populates a linked gallery of ERA5-derived input composites. }
  \label{fig:umap}
\end{figure}

\section{Demonstration: Tropical-Cyclone Retrieval}
Selecting an appropriate embedding model for a scientific retrieval task remains an open problem. Different models induce different notions of similarity, and the suitability of a representation depends on the specific question under investigation. The workbench is primarily designed to support this kind of exploration, enabling a researcher to apply a candidate model, examine whether its latent space exhibits structure aligned with known physical attributes, and assess its suitability for a given task. Accordingly, the following demonstration employs a vision model not to establish it as the recommended representation for meteorological retrieval, nor to benchmark its accuracy, but to illustrate how the platform supports the exploration of an embedding-based representation.

\begin{figure*}[t]
  \centering
  \includegraphics[width=\textwidth]{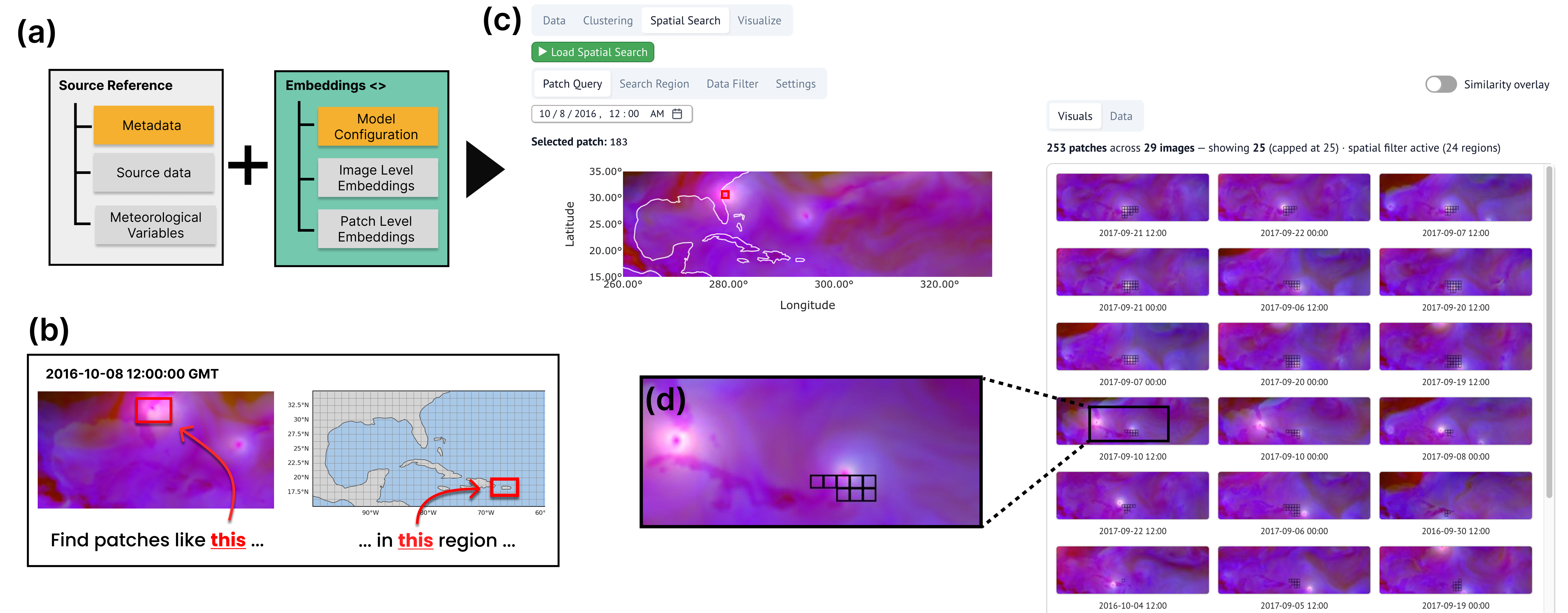}
  \caption{Visual-analogue search workflow: (a) linked source and embedding tables (See \cref{fig:teaser} (b))  serve as system inputs; (b) a conceptual schematic illustrates a localized patch query restricted to a geographic boundary; (c) the workbench interface executes this query via the Spatial Search tab and displays a gallery of retrieved analogs; and (d) a magnified inset highlights the specific matching patches within the results.}
  \label{fig:search}
\end{figure*}

\subsection{Dataset and Workflow Overview}

To illustrate an example workflow of how an analyst utilizing a vision-based model would interact with the workbench, we present a tropical-cyclone retrieval scenario. As a prerequisite to this workflow, image-level and patch-level embeddings are first generated for the entire dataset using the chosen model. This demonstration uses hourly ERA5 composites from the North Atlantic basin (2016 to 2018), paired with IBTrACS storm-track metadata. IBTrACS observations are spatiotemporally joined with each composite to provide storm labels and intensity metadata. The workflow described here illustrates the core retrieval loop; the interaction sequence, including additional views and features omitted for space, is shown in the supplemental screen recording.

\subsection{Input Encoding and Diagnostic Visualization}

Vision foundation models, particularly those based on contrastive learning (e.g., CLIP) and self-distillation (e.g., DINO), have been applied to Earth observation data and geographic information systems (GIS) \cite{brown_alphaearth_2025}. More recently, DINOv3~\cite{simeoni_dinov3_2025} has been shown to surpass prior remote sensing foundation models on tasks like open-vocabulary semantic segmentation without requiring domain-specific fine-tuning. We selected DINOv3 for this demonstration because it produces dense, patch-level features that can be used directly as a frozen backbone, without domain-specific fine-tuning.
To utilize the DINOv3 Base vision foundation model, which produces the required image-level and patch-level embeddings, ERA5 variables are encoded as three-channel images of mean sea-level pressure anomaly, 10 m wind speed, and total column water vapor (\cref{tab:rgb-encoding}; example composites in
\cref{fig:composite}a--c). This three-channel RGB mapping reflects the standard technical requirements of vision models, which expect typical image inputs.
The RGB mapping is a consequence of using a pretrained vision model. The rendered composites (\cref{fig:composite}) serve a diagnostic purpose, allowing the user to inspect exactly what the model sees, rather than acting as the primary visualization for human analytical reasoning. To support conventional meteorological analysis without the perceptual interference of mixed color channels, the workbench provides a separate view featuring univariate visualization of the underlying source data.  Representations learned directly from multi-variable meteorological data, such as those produced by multi-channel autoencoders~\cite{zhao_transforming_2025}, would remove this constraint, and the model-agnostic design of the workbench is intended to accommodate such encoders in the future.

\subsection{Query Construction and Patch-Based Search}
Before issuing a query, the user first inspects the DINOv3 latent space through a UMAP projection colored by IBTrACS maximum sustained wind speed (\cref{fig:umap}. Storm-containing composites form a distinct peripheral cluster. Box-selecting the cluster populates a linked gallery of raw composites, confirming that the structure is meteorological rather than a latent-space artifact. This step grounds the embedding experiment as suitable for retrieval.
The user then issues a localized query targeting a patch in Hurricane Matthew near the Florida coast (2016-10-08). To construct this query, the user selects a patch in the region of interest in the source image, which the system maps to the corresponding DINOv3 patch embeddings. The user then applies a spatial filter restricting matches to the Caribbean region around Puerto Rico (\cref{fig:search}). The backend executes this by applying a prefilter to the geospatial metadata index, followed by a nearest-neighbor vector search over the remaining valid patch embeddings from the pre-computed dataset. As expected, the nearest matches are Matthew’s own track frames followed by Hurricane Irma north of Puerto Rico (2017-09-08 to 2017-09-12) and Hurricane Maria during its landfall on the island (2017-09-15 to 2017-09-27). 

The primary contribution of this workbench is enabling the simultaneous search and visualization of both latent spaces (via embeddings) and physical state spaces (via meteorological variables). We do not attempt to quantify the accuracy of the retrieved results in this paper. Doing so meaningfully requires task-specific definitions of meteorological similarity, curated query sets, and interpretability of the embedding model - a study in its own right, and one for which the workbench is the supporting infrastructure rather than the object of evaluation. The workbench can isolate and export retrieval queries for exactly this kind of subsequent model-level benchmarking.

This scenario illustrates how the workbench supports a practical discovery workflow: a known phenomenon is used to characterize a region of interest in the embedding space, and a constrained patch-level query then surfaces meteorologically similar events across the archive. The same query construction can be reused against embedding experiments built over other corpora, such as AI-emulator ensembles, to search for similar phenomena where storm-track labels are not readily available.

\section{Scalability Evaluation}
\label{sec_scalability}
We benchmarked memory-mapped IVF-PQ indexes over patch-embedding corpora ranging from 0.98 Million(M) to 23.55 Million(M) 768-dimensional vectors. In a memory-mapped index, the data resides on disk , and the operating system pages only the needed portions into memory on demand. Benchmarks were run on an HPC compute node configured with 16 GiB RAM and local NVMe SSD staging, simulating a commodity workstation.

Ground-truth neighbors were computed by brute-force search. For each query, we exhaustively compared it against every vector in the full corpus and retained the exact top-10. This search used the raw float32 embeddings and the same cosine similarity metric as the index. This ground truth defines the baseline for Recall@10, our measure of approximate-search quality, reported as a percentage. Recall@10 tells us how much the speed optimizations degrade result accuracy relative to exact search, with values above roughly 90\% considered acceptable for exploratory retrieval and higher being better. Computing ground truth on the uncompressed vectors, rather than the quantized codes, ensures that Recall@10 measures the loss from approximate search alone.

For each index, we ran 100 untimed warm-up queries followed by 1000 timed queries. We report four metrics. Mean latency, in milliseconds, captures aggregate query cost, indicating whether queries return fast enough for interactive exploration (typically under 100 ms). Tail latency (p95/p99) reflects worst-case interactive behavior; keeping this tail near the mean is desirable since a low mean can hide occasional slow queries that disrupt sessions. Peak process Resident Set Size (RSS), in GiB, is the maximum physical memory the process occupies. It determines whether the index fits on commodity hardware, ideally staying well below available RAM.

\begin{figure}[tb]
 \centering 
 \includegraphics[width=\columnwidth]{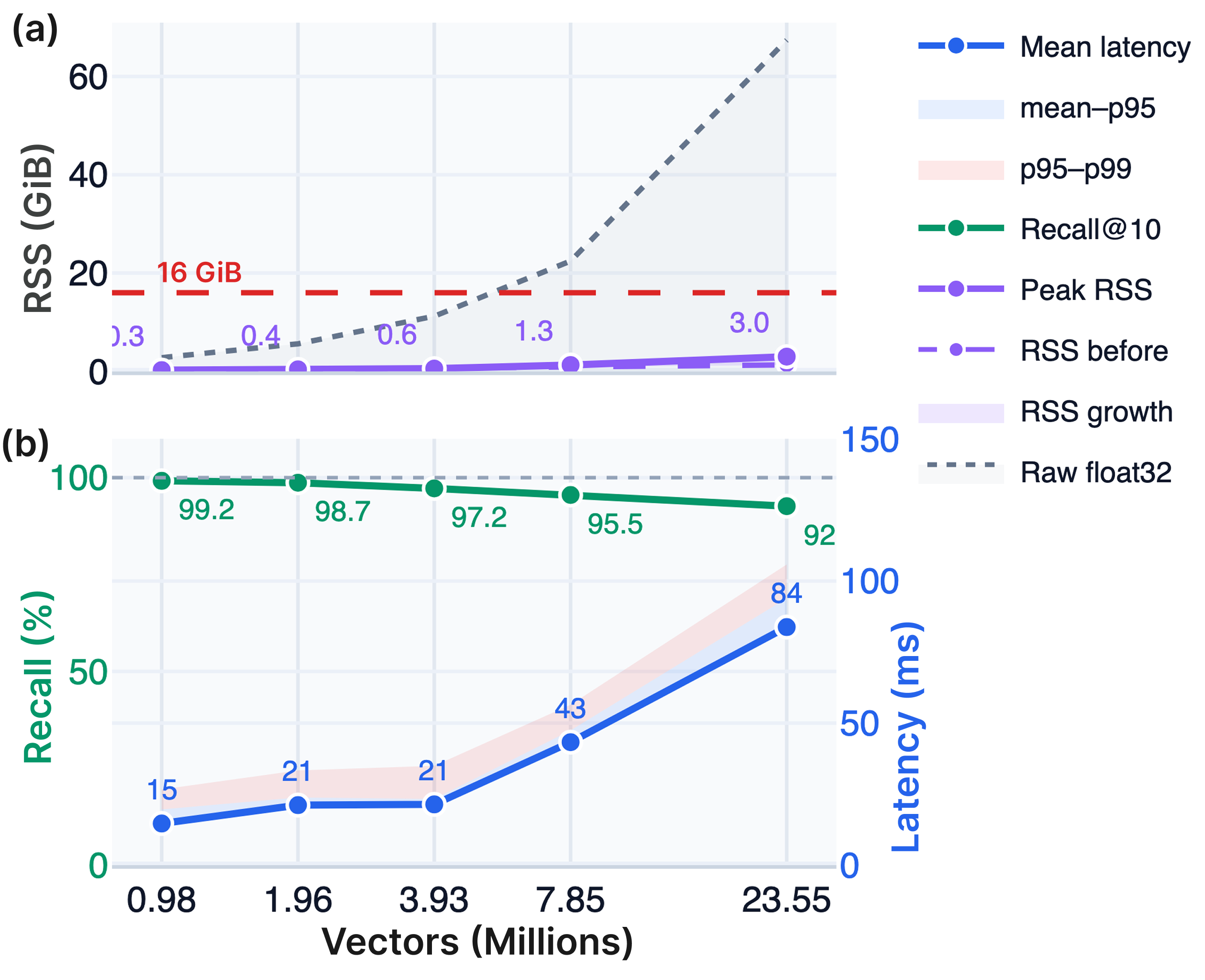}
 \caption{Scalability of out-of-core embedding retrieval. (a) As the index grows, process RSS scales sublinearly and remains far below the workstation's memory capacity. (b) Recall@10 stays above 92\% and mean latency remains under 100 ms.}
 \label{fig:scalability_eval}
\end{figure}

\cref{fig:scalability_eval} shows the backend supports out-of-core search well beyond available memory. The uncompressed float32 vector footprint exceeds the 16 GiB budget at larger scales, but peak RSS stays roughly at 3 GiB even for the 23.55M vector index. Since the full corpus is never loaded into process memory, there is substantial headroom for larger datasets on commodity hardware, bounded mainly by storage bandwidth and access patterns.

Latency rises with corpus size but stays practical for exploratory use, and Recall@10 remains high. Both can be traded against each other at query time, without rebuilding the index. This is controlled by two parameters: the number of IVF probes (a count of index partitions scanned per query) and the refinement depth (a count of candidates re-ranked using their fuller representations).

\section{Future Work}
We presented an open-source visual analytics workbench that addresses the growing challenge of exploring large, high-dimensional weather and climate datasets by treating embedding-based search as an inspectable, provenance-aware workflow. Organizing representations as embedding experiments tied to shared source data, metadata, and model configuration lets researchers examine how a latent space is organized, retrieve analogs through familiar meteorological views, and trace results back to the physical data that produced them. The tropical-cyclone demonstration illustrates this discovery loop, and the scalability evaluation shows that patch-level indexes of tens of millions of vectors can be searched out-of-core at interactive speed on commodity hardware. Recognizing the diversity of embedding models and retrieval strategies, and the rapid pace of representation learning, we designed the system as an extensible environment for experimentation rather than a single-purpose search interface.

This framing motivates two complementary axes of future work: what is searched (the representation) and how it is searched (the retrieval strategy); alongside continued collaboration with domain scientists to refine the user experience for query formulation and result interpretation.

\noindent\textbf{Embedding models.}
Although the demonstration used a general-purpose vision model with an RGB encoding, the embedding-experiment abstraction accommodates any model that produces fixed-length vectors linkable to source data, including autoencoders trained directly on multi-variable fields~\cite{zhao_transforming_2025}, domain-specific foundation models, and the latent spaces of AI weather emulators~\cite{richards_latent_2026}. Future work will use the workbench to systematically compare such representations, assessing whether each induces latent structure aligned with known physical attributes and how retrieval behavior shifts with preprocessing, spatial scale, and similarity definition. This qualitative auditing complements quantitative, model-level benchmarking: the workbench can isolate and export retrieval queries as reusable evaluation sets. We further plan to integrate these comparisons with explainability (XAI) workflows, so that differences between representations can be attributed to specific features, extending the attention-map inspection available in the current version.

\noindent\textbf{Search strategies.}
The current implementation supports image-level and patch-level similarity search with metadata and spatial filters. This is a foundation for substantially richer query constructions. Near-term extensions to the workbench include composing a query from multiple patch vectors, querying with vectors for segmented spatial regions, and multi-region queries. Beyond these immediate additions, because the system is designed as an open and extensible environment, it lays the groundwork for the community to explore progressively more ambitious retrieval paradigms.

Building on the system's core provenance links, researchers could adapt the architecture for compound event search by joining co-occurring features across regions and variables, for example, locating instances where a North American heatwave coincides with a Greenland blocking high. The framework similarly enables cross-dataset transfer, allowing users to formulate a query on a well-labeled reanalysis such as ERA5 and retrieve against a less-labeled corpus (e.g., matching an atmospheric river signature across CMIP6 historical runs) to probe how well embedding spaces align across datasets.

Furthermore, the system’s abstraction provides a foundation for historical-analogue search (retrieving dynamically similar past states from a named event, such as integrated vapor transport patterns analogous to Hurricane Harvey’s stall over southeast Texas) and precursor search (extending retrieval along the temporal axis to surface 5-10 day precursor patterns to states resembling a Pacific Northwest heat dome). Finally, through the integration of multi-modal models, the workbench could be extended to support impact-conditioned analogue search. This would condition retrieval on downstream consequences rather than atmospheric structure alone, for example, finding ensemble states whose track and intensity could produce wind-damage footprints in Florida comparable to Hurricane Irma. Together, these possibilities illustrate how the platform can span a broader research agenda: from single-to-compound queries, single-to-cross-dataset retrieval, spatial-to-spatiotemporal search, and feature-similarity-to-impact-conditioned matching.

\noindent\textbf{Language-driven interfaces.}
Finally, embeddings provide a natural bridge to multimodal and language-driven interaction with Earth system data. As large language and vision-language models become more integrated with scientific workflows~\cite{ding_bridging_2026}, future versions of the workbench could support richer query specification and cross-modal reasoning over data, metadata, and scientific literature.

\section*{Supplemental Materials}
\label{sec:supplemental_materials}

A screen recording demonstrating the workflow accompanies this paper. The source code can be accessed at \url{https://github.com/NCAR/scivis-embedding-workbench} which includes instructions and notebooks for generating the embeddings. Example embeddings for reproducing the tropical-cyclone demonstration and the scalability benchmarks is available through NSF NCAR's GDEX platform at \url{https://doi.org/10.5065/12ZJ-ZZ25}

\acknowledgments{
This work was supported by the National Center for Atmospheric Research, which is a major facility sponsored by the National Science Foundation under Cooperative Agreement 1852977.}

\bibliographystyle{abbrv-doi}

\bibliography{references}

@article{ding_bridging_2026,
	title = {Bridging data and discovery: a survey on knowledge graphs in {AI} for science},
	volume = {13},
	copyright = {https://creativecommons.org/licenses/by/4.0/},
	issn = {2095-5138, 2053-714X},
	shorttitle = {Bridging data and discovery},
	url = {https://academic.oup.com/nsr/article/doi/10.1093/nsr/nwag140/8507209},
	doi = {10.1093/nsr/nwag140},
	language = {en},
	number = {8},
	urldate = {2026-07-05},
	journal = {National Science Review},
	author = {Ding, Keyan and Zhu, Zhihui and Tang, Yuqi and Feng, Kehua and Zhuang, Xiang and Wang, Hongwei and Yang, Yi and Du, Huifang and Ni, Zhangkai and Wang, Shiqi and Fan, Xiaohui and Xing, Huabin and Bai, Lei and Liu, Qi and Wang, Haofen and Zhang, Qiang and Chen, Huajun},
	month = apr,
	year = {2026},
	pages = {nwag140},
}

@inproceedings{panta_web-based_2024,
	address = {St Pete Beach, FL, USA},
	title = {Web-based {Visualization} and {Analytics} of {Petascale} data: {Equity} as a {Tide} that {Lifts} {All} {Boats}},
	copyright = {https://doi.org/10.15223/policy-029},
	isbn = {979-8-3315-1692-5},
	shorttitle = {Web-based {Visualization} and {Analytics} of {Petascale} data},
	url = {https://ieeexplore.ieee.org/document/10767643/},
	doi = {10.1109/LDAV64567.2024.00009},
	urldate = {2026-05-01},
	booktitle = {2024 {IEEE} 14th {Symposium} on {Large} {Data} {Analysis} and {Visualization} ({LDAV})},
	publisher = {IEEE},
	author = {Panta, Aashish and Huang, Xuan and McCurdy, Nina and Ellsworth, David and Gooch, Amy A. and Scorzelli, Giorgio and Torres, Hector and Klein, Patrice and Ovando-Montejo, Gustavo A. and Pascucci, Valerio},
	month = oct,
	year = {2024},
	pages = {1--11},
}

@misc{tempest_mechanistic_2026,
	title = {Mechanistic {Interpretability} {Tool} for {AI} {Weather} {Models}},
	copyright = {Creative Commons Attribution 4.0 International},
	url = {https://arxiv.org/abs/2604.20467},
	doi = {10.48550/ARXIV.2604.20467},
	urldate = {2026-05-01},
	publisher = {arXiv},
	author = {Tempest, Kirsten I. and Beylich, Matthias and Craig, George C.},
	year = {2026},
	note = {Version Number: 1},
}

@article{prabhat_climatenet_2021,
	title = {{ClimateNet}: an expert-labeled open dataset and deep learning architecture for enabling high-precision analyses of extreme weather},
	volume = {14},
	issn = {1991-959X},
	shorttitle = {{ClimateNet}},
	url = {https://gmd.copernicus.org/articles/14/107/2021/},
	doi = {10.5194/gmd-14-107-2021},
	language = {English},
	number = {1},
	urldate = {2026-04-25},
	journal = {Geoscientific Model Development},
	publisher = {Copernicus GmbH},
	author = {Prabhat and Kashinath, Karthik and Mudigonda, Mayur and Kim, Sol and Kapp-Schwoerer, Lukas and Graubner, Andre and Karaismailoglu, Ege and von Kleist, Leo and Kurth, Thorsten and Greiner, Annette and Mahesh, Ankur and Yang, Kevin and Lewis, Colby and Chen, Jiayi and Lou, Andrew and Chandran, Sathyavat and Toms, Ben and Chapman, Will and Dagon, Katherine and Shields, Christine A. and O'Brien, Travis and Wehner, Michael and Collins, William},
	month = jan,
	year = {2021},
	pages = {107--124},
}

@incollection{keim_visual_2008,
	address = {Berlin, Heidelberg},
	title = {Visual {Analytics}: {Scope} and {Challenges}},
	isbn = {978-3-540-71080-6},
	shorttitle = {Visual {Analytics}},
	url = {https://doi.org/10.1007/978-3-540-71080-6_6},
	doi = {10.1007/978-3-540-71080-6_6},
	language = {en},
	urldate = {2026-04-25},
	booktitle = {Visual {Data} {Mining}: {Theory}, {Techniques} and {Tools} for {Visual} {Analytics}},
	publisher = {Springer},
	author = {Keim, Daniel A. and Mansmann, Florian and Schneidewind, Jörn and Thomas, Jim and Ziegler, Hartmut},
	editor = {Simoff, Simeon J. and Böhlen, Michael H. and Mazeika, Arturas},
	year = {2008},
	pages = {76--90},
}

@article{molina_review_2023,
	chapter = {Artificial Intelligence for the Earth Systems},
	title = {A {Review} of {Recent} and {Emerging} {Machine} {Learning} {Applications} for {Climate} {Variability} and {Weather} {Phenomena}},
	volume = {2},
	issn = {2769-7525},
	url = {https://journals.ametsoc.org/view/journals/aies/2/4/AIES-D-22-0086.1.xml},
	doi = {10.1175/AIES-D-22-0086.1},
	language = {EN},
	number = {4},
	urldate = {2026-04-25},
	journal = {Artificial Intelligence for the Earth Systems},
	publisher = {American Meteorological Society},
	author = {Molina, Maria J. and O’Brien, Travis A. and Anderson, Gemma and Ashfaq, Moetasim and Bennett, Katrina E. and Collins, William D. and Dagon, Katherine and Restrepo, Juan M. and Ullrich, Paul A.},
	month = sep,
	year = {2023},
}

@article{abernathey_cloud-native_2021,
	title = {Cloud-{Native} {Repositories} for {Big} {Scientific} {Data}},
	volume = {23},
	issn = {1558-366X},
	url = {https://ieeexplore.ieee.org/abstract/document/9354557},
	doi = {10.1109/MCSE.2021.3059437},
	number = {2},
	urldate = {2026-04-24},
	journal = {Computing in Science \& Engineering},
	author = {Abernathey, Ryan P. and Augspurger, Tom and Banihirwe, Anderson and Blackmon-Luca, Charles C. and Crone, Timothy J. and Gentemann, Chelle L. and Hamman, Joseph J. and Henderson, Naomi and Lepore, Chiara and McCaie, Theo A. and Robinson, Niall H. and Signell, Richard P.},
	month = mar,
	year = {2021},
	pages = {26--35},
}

@article{eyring_pushing_2024,
	title = {Pushing the frontiers in climate modelling and analysis with machine learning},
	volume = {14},
	issn = {1758-678X, 1758-6798},
	url = {https://www.nature.com/articles/s41558-024-02095-y},
	doi = {10.1038/s41558-024-02095-y},
	language = {en},
	number = {9},
	urldate = {2026-04-23},
	journal = {Nature Climate Change},
	author = {Eyring, Veronika and Collins, William D. and Gentine, Pierre and Barnes, Elizabeth A. and Barreiro, Marcelo and Beucler, Tom and Bocquet, Marc and Bretherton, Christopher S. and Christensen, Hannah M. and Dagon, Katherine and Gagne, David John and Hall, David and Hammerling, Dorit and Hoyer, Stephan and Iglesias-Suarez, Fernando and Lopez-Gomez, Ignacio and McGraw, Marie C. and Meehl, Gerald A. and Molina, Maria J. and Monteleoni, Claire and Mueller, Juliane and Pritchard, Michael S. and Rolnick, David and Runge, Jakob and Stier, Philip and Watt-Meyer, Oliver and Weigel, Katja and Yu, Rose and Zanna, Laure},
	month = sep,
	year = {2024},
	pages = {916--928},
}

@article{mahesh_huge_2025,
	title = {Huge ensembles – {Part} 2: {Properties} of a huge ensemble of hindcasts generated with spherical {Fourier} neural operators},
	volume = {18},
	issn = {1991-959X},
	shorttitle = {Huge ensembles – {Part} 2},
	url = {https://gmd.copernicus.org/articles/18/5605/2025/},
	doi = {10.5194/gmd-18-5605-2025},
	language = {English},
	number = {17},
	urldate = {2026-04-23},
	journal = {Geoscientific Model Development},
	publisher = {Copernicus GmbH},
	author = {Mahesh, Ankur and D. Collins, William and Bonev, Boris and Brenowitz, Noah and Cohen, Yair and Harrington, Peter and Kashinath, Karthik and Kurth, Thorsten and North, Joshua and O'Brien, Travis A. and Pritchard, Michael and Pruitt, David and Risser, Mark and Subramanian, Shashank and Willard, Jared},
	month = sep,
	year = {2025},
	pages = {5605--5633},
}

@article{bouallegue_rise_2024,
	chapter = {Bulletin of the American Meteorological Society},
	title = {The {Rise} of {Data}-{Driven} {Weather} {Forecasting}: {A} {First} {Statistical} {Assessment} of {Machine} {Learning}–{Based} {Weather} {Forecasts} in an {Operational}-{Like} {Context}},
	volume = {105},
	issn = {0003-0007, 1520-0477},
	shorttitle = {The {Rise} of {Data}-{Driven} {Weather} {Forecasting}},
	url = {https://journals.ametsoc.org/view/journals/bams/105/6/BAMS-D-23-0162.1.xml},
	doi = {10.1175/BAMS-D-23-0162.1},
	language = {EN},
	number = {6},
	urldate = {2026-04-23},
	journal = {Bulletin of the American Meteorological Society},
	publisher = {American Meteorological Society},
	author = {Bouallègue, Zied Ben and Clare, Mariana C. A. and Magnusson, Linus and Gascón, Estibaliz and Maier-Gerber, Michael and Janoušek, Martin and Rodwell, Mark and Pinault, Florian and Dramsch, Jesper S. and Lang, Simon T. K. and Raoult, Baudouin and Rabier, Florence and Chevallier, Matthieu and Sandu, Irina and Dueben, Peter and Chantry, Matthew and Pappenberger, Florian},
	month = jun,
	year = {2024},
	pages = {E864--E883},
}

@article{vance_big_2024,
	title = {Big data in {Earth} science: {Emerging} practice and promise},
	volume = {383},
	shorttitle = {Big data in {Earth} science},
	url = {https://www.science.org/doi/10.1126/science.adh9607},
	doi = {10.1126/science.adh9607},
	number = {6688},
	urldate = {2026-04-23},
	journal = {Science},
	publisher = {American Association for the Advancement of Science},
	author = {Vance, Tiffany C. and Huang, Thomas and Butler, Kevin A.},
	month = mar,
	year = {2024},
	pages = {eadh9607},
}

@article{overpeck_climate_2011,
	address = {New York, N.Y.},
	title = {Climate data challenges in the 21st century},
	volume = {331},
	issn = {1095-9203},
	doi = {10.1126/science.1197869},
	language = {eng},
	number = {6018},
	journal = {Science},
	author = {Overpeck, Jonathan T. and Meehl, Gerald A. and Bony, Sandrine and Easterling, David R.},
	month = feb,
	year = {2011},
	pages = {700--702},
}

@article{richards_latent_2026,
	title = {Latent {Representations} of {Land}–{Sea} {Boundaries} and {Extreme} {Temperature} in {Aurora}’s {Encoder} ({Student} {Abstract})},
	volume = {40},
	copyright = {Copyright (c) 2026 Association for the Advancement of Artificial Intelligence},
	issn = {2374-3468},
	url = {https://ojs.aaai.org/index.php/AAAI/article/view/42272},
	doi = {10.1609/aaai.v40i48.42272},
	language = {en},
	number = {48},
	urldate = {2026-04-23},
	journal = {Proceedings of the AAAI Conference on Artificial Intelligence},
	author = {Richards, Benjamin and Balan, Pushpa Kumar},
	month = mar,
	year = {2026},
	pages = {41368--41369},
}

@article{johnson_billion-scale_2021,
	title = {Billion-{Scale} {Similarity} {Search} with {GPUs}},
	volume = {7},
	issn = {2332-7790},
	url = {https://ieeexplore.ieee.org/document/8733051},
	doi = {10.1109/TBDATA.2019.2921572},
	number = {3},
	urldate = {2026-04-23},
	journal = {IEEE Transactions on Big Data},
	author = {Johnson, Jeff and Douze, Matthijs and Jégou, Hervé},
	month = jul,
	year = {2021},
	pages = {535--547},
}

@article{jegou_product_2011,
	title = {Product {Quantization} for {Nearest} {Neighbor} {Search}},
	volume = {33},
	copyright = {https://ieeexplore.ieee.org/Xplorehelp/downloads/license-information/IEEE.html},
	issn = {0162-8828},
	url = {http://ieeexplore.ieee.org/document/5432202/},
	doi = {10.1109/TPAMI.2010.57},
	language = {en},
	number = {1},
	urldate = {2026-04-23},
	journal = {IEEE Transactions on Pattern Analysis and Machine Intelligence},
	author = {Jégou, H and Douze, M and Schmid, C},
	month = jan,
	year = {2011},
	pages = {117--128},
}

@misc{pace_lance_2025,
	title = {Lance: {Efficient} {Random} {Access} in {Columnar} {Storage} through {Adaptive} {Structural} {Encodings}},
	shorttitle = {Lance},
	url = {http://arxiv.org/abs/2504.15247},
	doi = {10.48550/arXiv.2504.15247},
	urldate = {2026-04-20},
	publisher = {arXiv},
	author = {Pace, Weston and She, Chang and Xu, Lei and Jones, Will and Lockett, Albert and Wang, Jun and Shah, Raunak},
	month = apr,
	year = {2025},
	note = {arXiv:2504.15247 [cs]},
}

@article{kawakami_climatesom_2026,
	title = {{ClimateSOM}: {A} {Visual} {Analysis} {Workflow} for {Climate} {Ensemble} {Datasets}},
	volume = {32},
	issn = {1077-2626},
	shorttitle = {{ClimateSOM}},
	url = {https://www.computer.org/csdl/journal/tg/2026/01/11271132/2bZpLfT6Gfm},
	doi = {10.1109/TVCG.2025.3634788},
	language = {English},
	number = {01},
	urldate = {2026-04-19},
	journal = {IEEE Transactions on Visualization and Computer Graphics},
	publisher = {IEEE Computer Society},
	author = {Kawakami, Yuya and Cayan, Daniel and Liu, Dongyu and Ma, Kwan-Liu},
	month = jan,
	year = {2026},
	pages = {473--483},
}

@article{monahan_empirical_2009,
	chapter = {Journal of Climate},
	title = {Empirical {Orthogonal} {Functions}: {The} {Medium} is the {Message}},
	volume = {22},
	issn = {0894-8755, 1520-0442},
	shorttitle = {Empirical {Orthogonal} {Functions}},
	url = {https://journals.ametsoc.org/view/journals/clim/22/24/2009jcli3062.1.xml},
	doi = {10.1175/2009JCLI3062.1},
	language = {EN},
	number = {24},
	urldate = {2026-04-16},
	journal = {Journal of Climate},
	publisher = {American Meteorological Society},
	author = {Monahan, Adam H. and Fyfe, John C. and Ambaum, Maarten H. P. and Stephenson, David B. and North, Gerald R.},
	month = dec,
	year = {2009},
	pages = {6501--6514},
}

@misc{ren_embedding_2025,
	title = {Embedding {Atlas}: {Low}-{Friction}, {Interactive} {Embedding} {Visualization}},
	shorttitle = {Embedding {Atlas}},
	url = {http://arxiv.org/abs/2505.06386},
	doi = {10.48550/arXiv.2505.06386},
	urldate = {2026-04-16},
	publisher = {arXiv},
	author = {Ren, Donghao and Hohman, Fred and Lin, Halden and Moritz, Dominik},
	month = jul,
	year = {2025},
	note = {arXiv:2505.06386 [cs]},
}

@misc{zhao_transforming_2025,
	title = {Transforming {Weather} {Data} from {Pixel} to {Latent} {Space}},
	url = {http://arxiv.org/abs/2503.06623},
	doi = {10.48550/arXiv.2503.06623},
	urldate = {2026-04-16},
	publisher = {arXiv},
	author = {Zhao, Sijie and Liu, Feng and Zhang, Xueliang and Chen, Hao and Han, Tao and Gong, Junchao and Tao, Ran and Xiao, Pengfeng and Bai, Lei and Ouyang, Wanli},
	month = mar,
	year = {2025},
	note = {arXiv:2503.06623 [cs]},
}

@article{liu_representation_2025,
	title = {Representation learning for geospatial data},
	volume = {31},
	issn = {1947-5683},
	url = {https://doi.org/10.1080/19475683.2025.2552157},
	doi = {10.1080/19475683.2025.2552157},
	number = {4},
	urldate = {2026-04-16},
	journal = {Annals of GIS},
	publisher = {Taylor \& Francis},
	author = {Liu, Yu and Wang, Xuechen and Wang, Yidan and Huang, Fei and Huang, Yingjing and Li, Yong and Zhang, Weiyu and Gong, Shuhui and Mai, Gengchen and Yao, Yao and Yue, Yang and Li, Haifeng and Zhang, Fan},
	month = oct,
	year = {2025},
	note = {\_eprint: https://doi.org/10.1080/19475683.2025.2552157},
	pages = {557--583},
}

@article{sellars_grand_2018,
	chapter = {Bulletin of the American Meteorological Society},
	title = {“{Grand} {Challenges}” in {Big} {Data} and the {Earth} {Sciences}},
	volume = {99},
	issn = {0003-0007, 1520-0477},
	url = {https://journals.ametsoc.org/view/journals/bams/99/6/bams-d-17-0304.1.xml},
	doi = {10.1175/BAMS-D-17-0304.1},
	language = {EN},
	number = {6},
	urldate = {2026-04-16},
	journal = {Bulletin of the American Meteorological Society},
	publisher = {American Meteorological Society},
	author = {Sellars, S. L.},
	month = jun,
	year = {2018},
	pages = {ES95--ES98},
}

@misc{simeoni_dinov3_2025,
	title = {{DINOv3}},
	url = {http://arxiv.org/abs/2508.10104},
	doi = {10.48550/arXiv.2508.10104},
	urldate = {2026-04-15},
	publisher = {arXiv},
	author = {Siméoni, Oriane and Vo, Huy V. and Seitzer, Maximilian and Baldassarre, Federico and Oquab, Maxime and Jose, Cijo and Khalidov, Vasil and Szafraniec, Marc and Yi, Seungeun and Ramamonjisoa, Michaël and Massa, Francisco and Haziza, Daniel and Wehrstedt, Luca and Wang, Jianyuan and Darcet, Timothée and Moutakanni, Théo and Sentana, Leonel and Roberts, Claire and Vedaldi, Andrea and Tolan, Jamie and Brandt, John and Couprie, Camille and Mairal, Julien and Jégou, Hervé and Labatut, Patrick and Bojanowski, Piotr},
	month = aug,
	year = {2025},
	note = {arXiv:2508.10104 [cs]},
}

@misc{brown_alphaearth_2025,
	title = {{AlphaEarth} {Foundations}: {An} embedding field model for accurate and efficient global mapping from sparse label data},
	shorttitle = {{AlphaEarth} {Foundations}},
	url = {http://arxiv.org/abs/2507.22291},
	doi = {10.48550/arXiv.2507.22291},
	urldate = {2026-04-15},
	publisher = {arXiv},
	author = {Brown, Christopher F. and Kazmierski, Michal R. and Pasquarella, Valerie J. and Rucklidge, William J. and Samsikova, Masha and Zhang, Chenhui and Shelhamer, Evan and Lahera, Estefania and Wiles, Olivia and Ilyushchenko, Simon and Gorelick, Noel and Zhang, Lihui Lydia and Alj, Sophia and Schechter, Emily and Askay, Sean and Guinan, Oliver and Moore, Rebecca and Boukouvalas, Alexis and Kohli, Pushmeet},
	month = sep,
	year = {2025},
	note = {arXiv:2507.22291 [cs]},
}
\end{document}